# Transition Metal Chalcogenide Tin Sulfide Nanodimensional Films Align Liquid Crystals

*Asi Solodar*[*,1], *Ghadah AlZaidy*[2], *Chung-Che Huang*[2], *Daniel W. Hewak*[2], *Ibrahim Abdulhalim*[1,3]

[1]Ben Gurion University of the Negev, Department of Electro-Optics Engineering and The Ilse Katz Institute for Nanoscale Science and Technology, Beer Sheva 84105, Israel

[2]Optoelectronics Research Centre, University of Southampton, Southampton, SO17 1BJ, U.K.

[3]Singapore-HUJ Alliance for Research and Enterprise (SHARE), Nanomaterials for Energy and Energy-Water Nexus (NEW), Campus for Research Excellence and Technological Enterprise (CREATE), Singapore 138602

ABSTRACT

Transition metal chalcogenide tin sulfide (SnS) films as alternative noncontact alignment layer for liquid crystals, have been demonstrated and investigated. The SnS has an anisotropic atomic chain structure similar to black Phosphorous which causes the liquid crystal molecules to align without the need for any additional surface treatments. The high anisotropic nature of SnS promotes the alignment of the easy axis of liquid crystal molecules along the periodic atomic grooves of the SnS layer. The atomically thin SnS layers were deposited on indium tin oxide films on glass substrates, at room temperature by chemical vapor deposition. The device characteristics are comparable to those commercially available, which use photo-aligning polymer materials. We measured threshold voltage of 0.92V, anchoring energy of $1.573 \times 10^{-6} \ J/m^2$, contrast ratio better than 71:1 and electro-optical rise/fall times of 80/390ms, respectively for ~11 micron thick liquid crystal device as expected.

1. **INTRODUCTION**

Over the past several years, 2D materials with diverse properties are under investigation for their great potential in photonic and electronic applications. Examples of such newly rising materials include Graphene, transition metal (di)-chalcogenides and black-phoshorous.[1] The bulk black phosphorus (BP) is the most stable allotrope of the phosphorus and has a high hole mobility, strong in-plane anisotropy and tunable direct bandgap energy of about ~0.3 eV to ~2 eV, covering the visible to infrared spectrum.[2,3] Recently, alignment of rod-like LC molecules, such as thermotropic and lyotropic LCs on black phosphorus surfaces were demonstrated.[4] However, the preparation of BP is complicated and involves high temperature and pressure. In addition the BP films are very sensitive to water vapor and oxygen, which leads to their rapid degradation, when exposed to air .[5,6] On the other hand, the study of Chao et al.,[7] assessed the Tin sulfide (SnS) material as a compound analogue of elemental black phosphorus with higher chemical stability than the latter one. The SnS material is a IV-VI compound semiconductor with direct band gap energy 1.2 eV to ~1.5 eV, high carrier mobility and high in-plane anisotropy.[8] In addition SnS exhibits a chemical inert surface without dangling bonds and surface density states, which makes it chemically and environmentally stable. Furthermore, the mere fact of high in-plane anisotropy, can teach on possible alignment abilities of LC molecules.

Liquid crystals (LC) are of a great importance for many electro-optical applications. They have found wide use in a large variety of electro-optical modules, such as flat panel displays,[9] linear polarization rotators,[10] dynamical wave plate retarders,[11] spatial light modulators and tunable filters. [12-15] A typical LC device has a configuration of a sandwich cell in which LC is filled between a pair of glass substrates coated with thin alignment layer on top of an indium tin oxide (ITO) films, used as transparent conducting electrodes. The alignment layers determine the orientation of LC molecules on the surface in the absence of an external electric field, and therefore the quality of the device and its response to an applied electric field. As a result, the interaction between the LC molecules and the alignment layer is of great importance in determining the device performance. Obtaining high contrast LC devices is immensely important for certain imaging and photonic applications; hence better alignment layers that can improve the devices quality is still in demand. In addition the fundamental science behind the interfacial interactions of liquid crystals and alignment layers is of high interest.

Several surface treatment methods have been developed during the last century to form the surface alignment films for LCs. Such well known alignment techniques as unidirectional mechanical rubbing of thin polymer or polyimide spin coated on glass substrate,[16] photo-alignment by illumination with ultraviolet polarized light of photosensitive polymers and chemically by surface active agents (surfactants).[17] As a result, LC molecules are aligned through the surface anisotropy caused by either the micro/nano-grooves, which had been formed on the rubbed surface, or by the interaction with polymer arranged molecular chains in photo activated polymers or interaction with vertically aligned surfactant molecules on the surface. Although widely used in the industry, the mechanical rubbing method has some shortcomings, as it involves physical contact interaction between the cloth and the surface, which can introduce particle dust generation and surface damage. On the other hand, noncontact methods such as photo–alignment, suffer from insufficient time stability and lower anchoring energy.

In order to overcome the imperfections of traditional aforementioned techniques, some alternative surface alignment methods have been investigated based on noncontact ideology. New concepts for photo alignment were investigated by our group, where nano dimensional chalcogenide glass films evaporated in vacuum and utilized for LC alignment as well as photo-sensing film for optically addressed spatial light modulators.[18,19] Analogous to conventional photoalignment materials, this method exhibits photo-induced anisotropy, stimulated by polarized blue light irradiation (436 nm). Another method was proposed by Kitson et al.,[20] where they demonstrated that tilted micron-scale posts prepared by photolithographic processes, can be used to generate uniform liquid crystal alignment. In addition, our group recently reported a direct writing of nano structured patterns on ITO coated glass surface by ultrafast femtosecond laser and LC alignment on lead sulphide (PbS) nano-sculptured thin films with glancing angle deposition technique as an alternative method for LC alignment layers with different alignment orientations and better transparency.[21-24] Apart from photo-induced optical anisotropy behavior, other types of alignment are manifested as an artificial generation of anisotropy on top of the surface and are not a result of intermolecular bonds. Therefore, those alignment methods at this stage cannot guarantee that LC molecules will orient unequivocally in preferable alignment direction and thus, at the same time can exhibit anti-parallel and parallel configurations as well.

In this work, we report on a novel alignment method of LC molecules based on the anisotropic nature of the SnS crystal structure without any additional surface treatments. The atomically thin SnS layers were deposited on indium tin oxide (ITO) films on top of the glass substrates, at room temperature by chemical vapor deposition (CVD) techniques. High contrast LC devices are shown to be achievable with this methodology.

## 2. EXPERIMENTAL SECTION

In this work, we have used two ITO coated glass substrates, each of dimensions 17x15x2.5 mm$^3$ coated with 145$\pm$10 nm ITO film (purchased from Thin Film Devices Inc. California, USA) and assembled on either side of the LC device. Prior to use, the substrates were cleaned by ultrasonic bath with deionized (DI) water, isopropanol (IPA) and at last with acetone, where each cycle lasts about 20 minutes at a temperature of 50 °C.

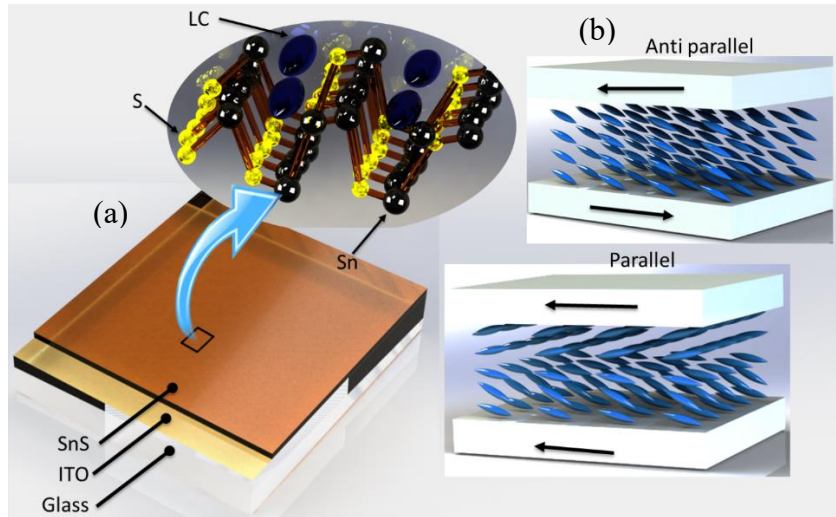

**Figure 1.** LC device assembly. (a) Deposition of SnS layer on top of the ITO coated glass substrate and SnS molecular structure with LC molecules aligned along the periodic atomic grooves. (b) Illustration of two possible alignment configurations, which can be achieved by this method.

Hereafter, SnS thin films were deposited on the ITO substrates (Figure 1a) by CVD with $SnCl_4$ precursor to react with $H_2S$ gas at room temperature and then annealed at 450 °C for 2 hours.[25] Then, the liquid crystal mixture E-7 (Merck) was sandwiched between SnS coated substrates, separated by 10 micron spacers. Generally, at this stage two possible configurations of anti-parallel and parallel alignment modes can be achieved as depicted in Figure 1b.

The extinction and alignment states of the devices were examined under polarization microscope (Olympus mx50) between crossed polarizers in reflection mode. The linearly polarized light (P) was focused by x5 objective onto the device as demonstrated in Figure 2. Upon reflection from the surface, the linearly polarized light beam is then split by the beam splitter, while part of it is passing through the analyzer (A), which is perpendicularly oriented with respect to polarizer axis.

First, the liquid crystal device was placed with its optical axis parallel to the axis of one of the polarizers in order to examine the extinction level. Then the LC alignment axis was oriented at 45º towards the polarizer axis, which will provide maximum reflected light intensity.

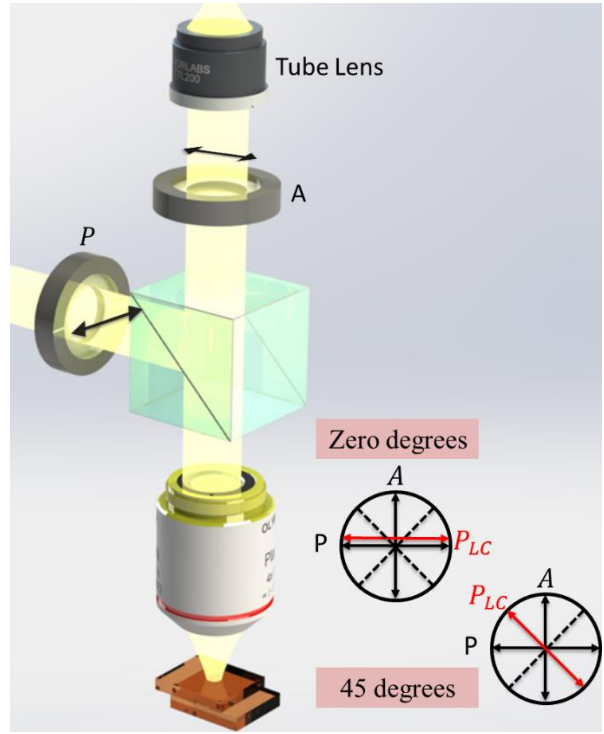

**Figure 2.** Illustration of LC device between crossed polarizer (P) and analyzer (A), under the polarization microscope in reflection operational mode. Demonstration of two working configurations: (1) Extinction (zero degrees) configuration with LC alignment axis ($P_{LC}$) being parallel to the polarizer axis and (2) maximum reflected light intensity, where $P_{LC}$ axis oriented at $45^0$ with respect to the polarizers axis.

The transmission spectrum characteristics of the device were studied by the optical setup, illustrated in Figure 3. The device was placed at 45º between crossed polarizers, with respect to the polarizer axis. White collimated linearly polarized light passed through the LC device and focused onto the fiber spectrometer by the x5 objective.

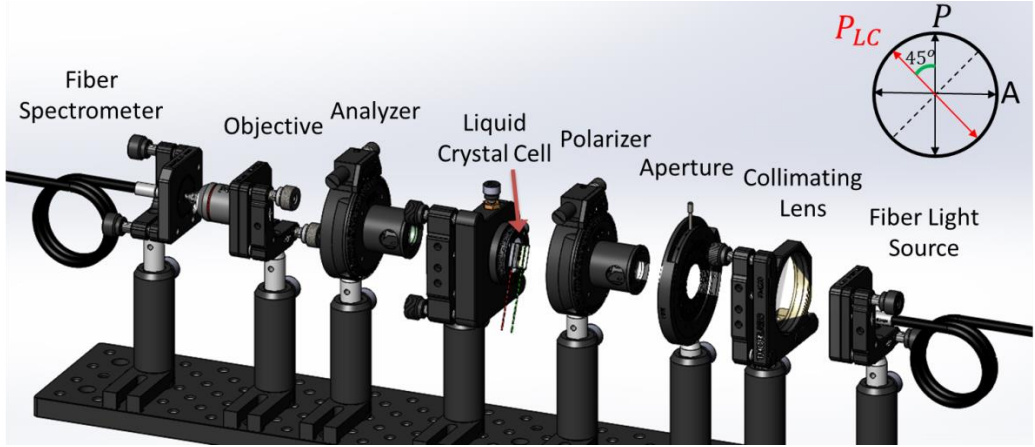

**Figure 3.** Schematic representation of the electro optical setup to determine transmission spectrum of the LC device. The setup comprises a fiber light source collimated using a lens with focal length f=8cm, polarizer, aperture, the LC device attached to the 6-axis kinematic mount, an analyzer, a x5 focusing microscope objective, an output fiber connected to spectrometer (StellarNet Inc.).

## 3. RESULTS AND DISCUSSION

First, in order to illustrate the LC molecules alignment along the periodic grooves of SnS layer, we used polarizing microscope as illustrated in Figure 2. The device placed between crossed polarizers and the alignment direction (top surface easy axis) oriented parallel to the polarizer axis as illustrated in zero degrees configuration in Figure 2. As a result, a high extinction level was achieved as depicted in the inset of Figure 4. It is clearly evident that the LC molecular alignment axis (easy axis) coincident with the SnS molecular grooves direction. Then, the device was oriented 45° with respect to the polarizers as illustrated in the 45° configuration in Figure 2, A 2kHz sinusoidal wave with variable amplitude voltage values was applied across the device in order to reorient the liquid crystal director.

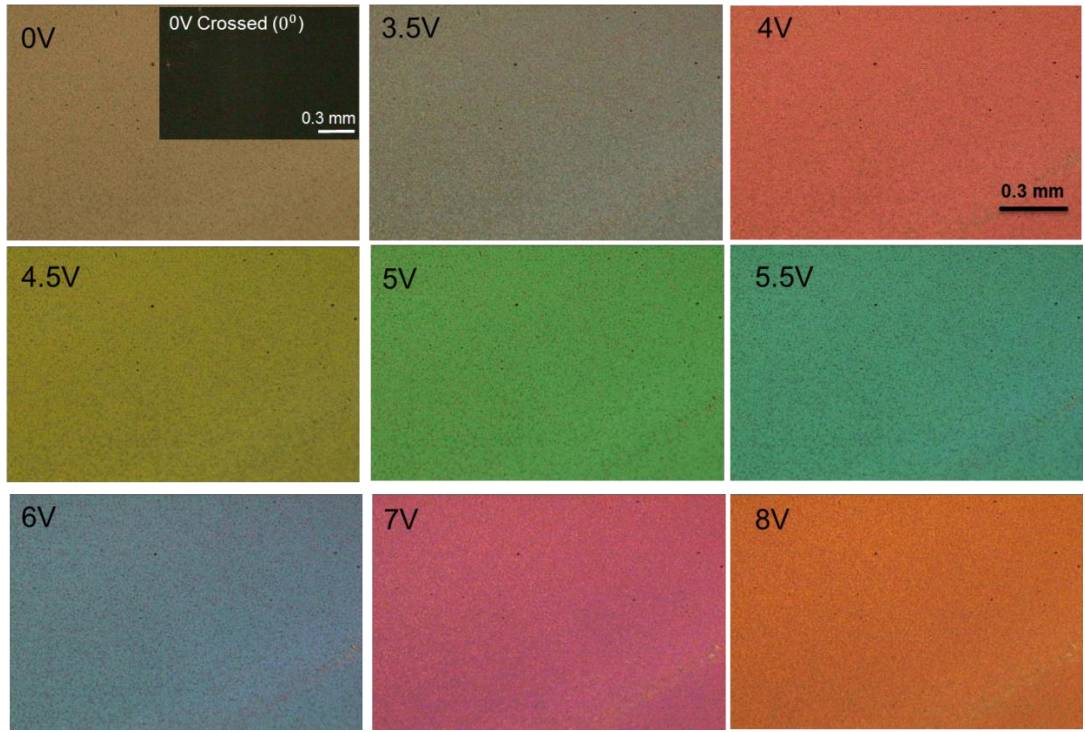

**Figure 4.** Camera read out of LC reflection intensities under polarized microscope between crossed polarizers as illustrated in Figure 2. The LC device rotated at 45 degrees with respect to the polarizer axis, while various amplitude AC voltages were applied. Inset black image in the 0V case shows extinction position when the LC easy axis oriented parallel to the axis of the polarizer.

As demonstrated in Figure 4 the device exhibits a uniform reflection change and tunable color at various bias voltages. In addition, a uniform extinction is demonstrated, when the LC device alignment axis is parallel to the axis of one of the polarizers as depicted in Figure 2a.

Characterization of the LC device as a function of voltage was achieved quantitatively, also by spectral transmission measurements. The device was placed between crossed polarizers at an angle of $45°$ between the easy axis of the LC ($P_{LC}$) and the polarizer as depicted in Figure 3. The measured transmission spectra for various applied voltages are shown in Figure 5.

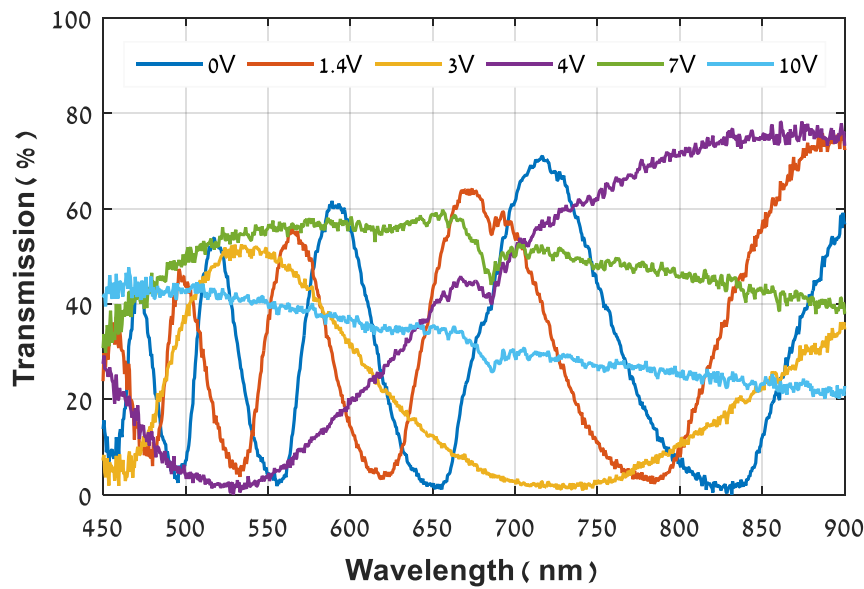

**Figure 5.** Spectrometer readout from electro optical setup demonstrated in Figure 3. Transmission spectra at different applied voltages (sinusoidal wave at 1 kHz) between crossed polarizers.

Upon reaching the threshold voltage, the transmission spectrum considerably changed (Figure 5). As can be seen, the transmission spectrum can be effectively controlled by the applied voltage, which changes the orientation of the LC director. As expected, for high enough bias voltages the alignment in the bulk is practically parallel to the field, thus the incoming linear polarized light feels a very weak optical anisotropy, resulting from residual influence of fixed boundary conditions. Thus, any further voltage increase will reduce the transmission until it will be completely blocked by the analyzer.

In order to obtain maximum accuracy during further electro optical measurements, a photodetector was used conjugated to lock-in amplifier SR830 (Stanford Research Systems) as illustrated in Figure 6a. The LC device was placed at 45° between crossed polarizers in order to receive maximum transmission, while a He-Ne laser (633 nm) was used as a light source.

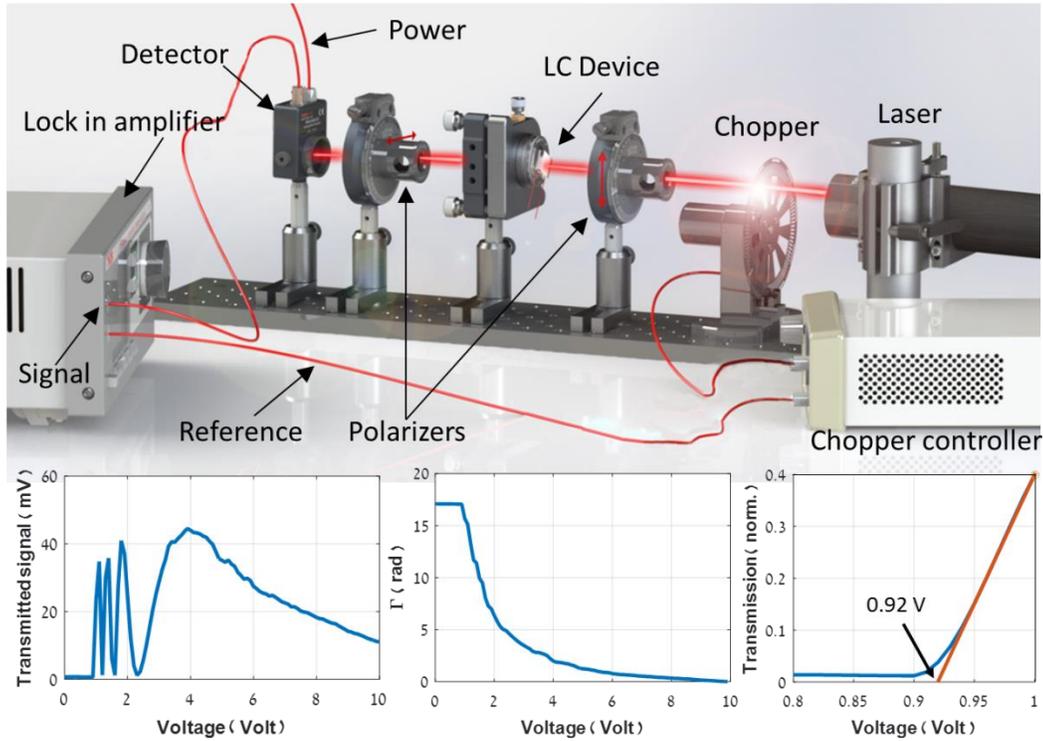

**Figure 6.** Illustration of electro optical setup for transmission measurement through LC device (a). Modulated by the chopper HeNe laser source was transmitted through the LC device, which was placed at $45^0$ between crossed polarizers. (b) Transmitted signal detected by the photo-detector. (c) Accumulated retardation of the device, deduced from the transmitted signal. (d) Measured threshold voltage.

The applied voltage to the device was swept in steps of 0.1V, while the transmission signal was detected by a photodiode connected to the lock-in amplifier. Based on the transmitted intensity signal (Figure 6b), we can deduce the accumulated phase retardation of the device as a function of voltage. As shown in Figure 6c, the retardation tends to decrease at higher applied voltages as expected for homogeneously aligned nematic LCs with positive dielectric anisotropy.

In addition, we can also utilize the transmitted signal obtained by the electro optical setup shown in Figure 6, in order to determine the threshold voltage of the LC device. As can be seen from the linear fit in Figure 6d, the measured threshold voltage is 0.92V, which is in very close agreement to the value calculated based on the equation for splayed cell:

$$V_{th}^{splay} = \pi \sqrt{\frac{K_{11}}{\varepsilon_0 \Delta \varepsilon}} \tag{1}$$

Here $K_{11}$ is a splay elastic constant, $\Delta \varepsilon$ is a dielectric anisotropy and $\varepsilon_0$ is the vacuum permittivity. For LC E7: $\Delta \varepsilon = 14$, $K_{11} = 10.8\ pN$.[26]

In order to measure the switching speed of the device, a modulated AC signal was applied. The light from a collimated laser diode (633nm) transmitted through the device and detected by a photodiode, which is connected to a digitizing oscilloscope. The switching speed is defined as the time required for a change from 10% to 90% of the total transmission change. The rise and fall times of the LC were approximately 80ms and 390ms, respectively as shown in Figure 7.

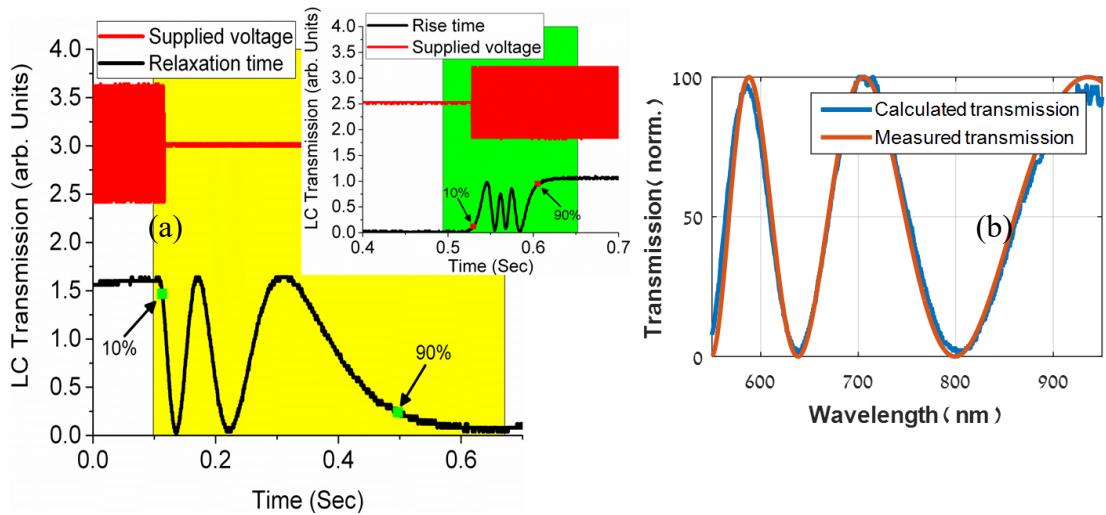

**Figure 7.** Measured electro-optical time response of the LC device (a). The transient oscillations are a result of the relatively thick LC cell. The measured decay time is about 390ms and the rise time is around 80ms. (b) Measured and simulated transmission signal for normal incident light through the LC device, between crossed polarizers, oriented at 45-degree.

The relaxation time ($t_{relax}$) and the rise time ($t_{rise}$) of a LC device can be evaluated by the dynamics of the Freedericksz transition under the one splay elastic constant approximation and can be calculated by:

$$t_{relax} = \frac{\gamma_\theta}{K_{11}}\left(\frac{d}{\pi}\right)^2, \quad t_{rise} = \frac{t_{relax}}{(V/V_c^{splay})^2 - 1} \qquad (2)$$

Here $\gamma_\theta$ is the rotational viscosity of the LC, d is the thickness of the LC layer and V is the RMS value of an applied voltage supply. In order to calculate the theoretical value of the response time we need an accurate measurement of the thickness. The thickness of the cell was determined by fitting the experimentally measured transmission spectrum through the device for normal incidence light to the theoretical simulated results (Figure 7b), given by Equation 3:

$$T = cos^2(P - A) - \sin(2P) \cdot \sin(2A) \cdot sin^2(\Gamma_{LC}/2) \qquad (3)$$

Here P and A are the angles between the LC device alignment axis and the polarizer / analyzer, respectively and $\Gamma$ is the LC phase retardation, defined as:

$$\Gamma_{LC} = \frac{2\pi}{\lambda} \int_0^d (n_{eff}(z) - n_o)dz \qquad (4)$$

Here $\lambda$ is the wavelength, z is the depths coordinate normal to the cell, $n_o$ and $n_{eff}(z)$ are the ordinary and the extraordinary refractive index of the LC, respectively and given by:

$$n_{eff}(z) = \frac{n_\perp n_\parallel}{\sqrt{n_\perp^2 cos^2\theta(z) + n_\parallel^2 sin^2\theta(z)}}, \quad n_o = n_\perp \qquad (5)$$

For LC E-7: $n_\parallel = 1.693$ and $n_\perp = 1.499$.

The extraordinary refractive index depends on the LC angle $\theta(z)$ and in the absence of external electric field can be interpreted as the LC pre-tilt angle. As apparent from Equations 2-5 and Figure 7b, the thickness of LC device is $d = 11.52 \mu m$ with a pre-tilt angle of $\theta(0,d) = 2°$. Thus, according to Equation 2, the fall and rise times of the LC should be about 302ms and 11ms (for V=5V), respectively. Despite the deviation between the measured and calculated switching times, the results are still in a plausible range of thick time response LC devices. Part of the disagreement is due to the fact that the expressions in equation 2 are approximations for the small deformations, very small pretilt angle and strong anchoring cases while here we used 5V which is more than five folds the threshold voltage and the anchoring as will be shown below can be considered intermediate. Another reason is the fact that we are measuring the electrooptic response times while the expressions in equation 2 refer to the molecules rotational motion. For example the relaxation time under the intermediate anchoring condition can be written as[22]:

$$\tau'_{relax} = \tau_{relax}\left(1 + \frac{4K}{dW}\right) \qquad (6)$$

Using the values of the elastic constant, the thickness and the anchoring energy of $1.573 \times 10^{-6}$ $J/m^2$ as estimated below we get that the relaxation time when considering the anchoring energy effect should be larger by a factor of x3 bringing it to around 900 msec and the rise time to around 30msec. The measure fall and rise times are 390 msec and 80 msec. The relaxation time is in good agreement if we consider the fact that the electrooptical relaxation time is nearly half the molecular relaxation time.[27] The rise time measured electro-optically can be larger than the molecular rise time depending on the applied voltage. An increase by factor of x2.5 can be explained as originating from the fact that the pretilt angle is 2 degrees, the use of large deformation and the effect of the electrooptical transfer function.[27]

As mentioned slower switching times may also indicate intermediate surface anchoring energy and defect domains. However; no defects observed over the entire active area, under the polarized microscope with and without applied voltages. The strength of the anchoring ($W_A$) between the nematic LC molecules and the SnS alignment layer can be determined by improved high electric field method, proposed by Nastishin et al.[28] This method is based on linear fit of the voltage $(V - V')$ and phase retardation normalized by the zero voltage phase $\Gamma_{LC}(V - \bar{V})/\Gamma_{LC}^0$. Here V is an applied voltage and $V' = \alpha(\Delta\varepsilon/\varepsilon_\parallel)V_c^{splay}$, where $\alpha$ coefficient varies between $1 - 2/\pi$. Additional discussions and various aspects of this method are described elsewhere.[28] The value of measured anchoring energy was on the order of $1.573 \times 10^{-6} \, J/m^2$ as deduced from Figure 8.

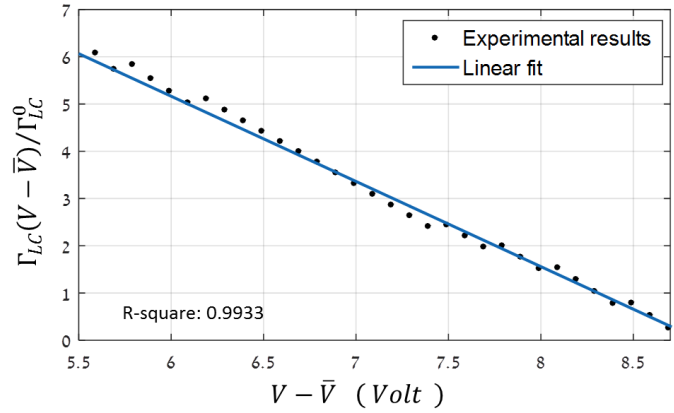

**Figure 8.** Plot of $\Gamma_{LC}(V - \bar{V})/\Gamma_{LC}^o$ as function of $(V - \bar{V})$ and linear fit used for anchoring energy evaluation ($W_A$), measured by electric field techniques,[28] yielding the value of $1.573 \times 10^{-6} \, J/m^2$.

This anchoring energy is of the same magnitude range as that measured using commercial photo-aligning polymer materials, which is usually limited to between $10^{-5} - 10^{-6} \, Jm^{-2}$.[29,30]

Another important measure of the performance of a LC based devices is the contrast ratio. From measurement of the transmitted intensity signal in Figure 6b, we found the contrast ratio of: $T_{max}/T_{min}$ =44.487/0.622 = 71: 1. We believe that this value can be increased after further optimization process such as better cleaned surfaces prior to SnS coating.

## 4. CONCLUSIONS

The control of molecular alignment in liquid-crystal based applications has a high importance in the overall performance of the devices, thus new alignment techniques are still extensively investigated. Here we demonstrated an alternative noncontact alignment method based on the naturally highly anisotropic tin sulfide crystal structure.

The molecular characteristics of di-chalcogenide SnS material is very similar to the structure of black phosphors, however in the former case the material is much more stable against atmospheric interactions. The LC molecules clearly reveal the preferred orientation along the molecular chains of the SnS films and successfully reorients as a result of electrical supplied voltage. The device exhibits threshold voltage of 0.92V, which is typical to nematic phase LC devices. In addition the measured anchoring energy is about $1.573 \times 10^{-6} \, J/m^2$ comparable with the anchoring energy of nematic LC on photosensitive polymers. The rise and fall switching times of 80ms/390ms, respectively also demonstrate plausible switching times for thick LC devices. The contrast ratio at this stage is about 71:1, but can be increased by optimization process. We believe that this alignment technique is of high potential due to its simplicity, limited in scale by only the size of the CVD chamber and the high contrast that can be achieved.

**Corresponding Author**

*(IL). Tel: (+972)08-6428599  E-mail: asisolodar@gmail.com

**ACKNOWLEDGMENT**

This research is partially supported by grants from the National Research Foundation, Prime Minister's Office, Singapore under its Campus of Research Excellence and Technological Enterprise (CREATE) programme. The 2D materials work at the University of Southampton is funded in part through the Future Photonics Manufacturing Hub (EPSRC EP/N00762X/1).

# REFERENCES

[1] X. Ling, H. Wang, S. X. Huang, F. N. Xia, M. S. Dresselhaus, The renaissance of black phosphorus. P Natl Acad Sci USA 2015, 112 (15), 4523-4530.

[2] F. N. Xia, H.Wang, Y. C. Jia, Rediscovering black phosphorus as an anisotropic layered material for optoelectronics and electronics. Nat Commun 2014, 5, 4458.

[3] H. Liu, Y. C. Du, Y. X. Deng, P. D. Ye, Semiconducting black phosphorus: synthesis, transport properties and electronic applications. Chem Soc Rev 2015, 44 (9), 2732-2743.

[4] D. W. Kim, H. S. Jeong, K. O. Kwon, J. M. Ok, S. M. Kim, H. T. Jung, Ultrastrong Anchoring on the Periodic Atomic Grooves of Black Phosphorus. Adv Mater Interfaces 2016, 3 (22), 1600534.

[5] J. O. Island, G. A. Steele, H. S. J. van der Zant, A. C. Gomez, Environmental instability of few-layer black phosphorus. 2d Mater 2015, 2 (1), 011002.

[6] A. Ziletti, A. Carvalho, D. K. Campbell, D. F. Coker, A. H. C. Neto, Oxygen Defects in Phosphorene. Phys Rev Lett 2015, 114 (4), 046801.

[7] C. Xin, J. X. Zheng, Y. T. Su, S. K. Li, B. K. Zhang, Y. C. Feng, F. Pan, Few-Layer Tin Sulfide: A New Black-Phosphorus-Analogue 2D Material with a Sizeable Band Gap, Odd-Even Quantum Confinement Effect, and High Carrier Mobility. J Phys Chem C 2016, 120 (39), 22663-22669.

[8] N. K. Reddy, M. Devika, E. S. R. Gopal, Review on Tin (II) Sulfide (SnS) Material: Synthesis, Properties, and Applications. Crit Rev Solid State 2015, 40 (6), 359-398.

[9] Khoo, I.-C. Liquid crystals: physical properties and nonlinear optical phenomena, Wiley: New York, N.Y., 1995.

[10] I. Abdulhalim, Liquid crystal active nanophotonics and plasmonics: from science to devices. J Nanophotonics 2012, 6, 061001.

[11] D.-K. Yang, S.-T. Wu, Fundamentals of liquid crystal devices, 2 ed.; John Wiley: Chichester ; Hoboken, NJ, 2006.

[12] P. K. Shrestha, Y. T. Chun, D. P. Chu, high-resolution optically addressed spatial light modulator based on ZnO nanoparticles. Light-Sci Appl 2015, 4,e259.

[13] A. Solodar, T. A. Kumar, G. Sarusi, I. Abdulhalim, Infrared to visible image up-conversion using optically addressed spatial light modulator utilizing liquid crystal and InGaAs photodiodes. Appl Phys Lett 2016, 108 (2), 021103.

[14] S. T. Wu, Design of a Liquid-Crystal Based Tunable Electrooptic Filter. Appl Optics 1989, 28 (1), 48-52.

[15] L. Vicari, Optical applications of liquid crystals, Institute of Physics Pub.: Bristol, 2003.

[16] J. M. Geary, J. W. Goodby, A. R. Kmetz, J. S. Patel, The Mechanism of Polymer Alignment of Liquid-Crystal Materials. J Appl Phys 1987, 62 (10), 4100-4108.

[17] N. Kawatsuki, H. Ono, H. Takatsuka, T. Yamamoto, O. Sangen, Liquid crystal alignment on photoreactive side-chain liquid-crystalline polymer generated by linearly polarized UV light. Macromolecules 1997, 30 (21), 6680-6682.

[18] I. Abdulhalim, M. G. Kirzhner, Y. Kurioz, M. Klebanov, V. Lyubin, Y. Reznikov, N. Sheremet, K. Slyusarenko, Integration of chalcogenide glassy films and liquid crystals for photoalignment and optically addressed modulators. Phys Status Solidi B 2012, 249 (10), 2040-2046.

[19] M. Gelbaor, M. Klebanov, V. Lyubin, I. Abdulhalim, Permanent photoalignment of liquid crystals on nanostructured chalcogenide glassy thin films. Appl Phys Lett 2011, 98 (7), 071909.

[20] S. C. Kitson, E. G. Edwards, A. D. Geisow, Designing liquid crystal alignment surfaces. Appl Phys Lett 2008, 92 (7), 073503.

[21] A. Solodar, A. Cerkauskaite, R. Drevinskas, P. G. Kazansky, I. Abdulhalim, Ultrafast Laser Nanostructured ITO Acts as Liquid Crystal Alignment Layer and Higher Transparency Electrode. To be published.

[22] A. Chaudhary, M. Klebanov, I. Abdulhalim, Liquid crystals alignment with PbS nanosculptured thin films. Liquid Crystals 2017, 1-8.

[23] A. Solodar, A. Cerkauskaite, R. Drevinskas, P. G. Kazansky, I. Abdulhalim, Liquid crystal alignment on ultrafast laser nanostructured ITO coated glass, CLEO/Europe-EQEC Munich, 2017. https://eprints.soton.ac.uk/410668/

[24] A. Solodar, A. Cerkauskaite, R. Drevinskas, P. G. Kazansky, I. Abdulhalim, Nanogratings Written by a Femto-second Laser on ITO Act as Higher Transparency Electrode and Alignment Layer for Liquid Crystals, ICMAT, Singapore, 2017. https://www.conftool.pro/icmat2017/index.php?page=browseSessions&form_session=351

[25] G. Alzaidy, C. C. Huang, D. W. Hewak, CVD-grown tin sulphide for thin film solar cell devices. In 11th Photovoltaic Science Applications and Technology (PVSAT-11), United Kingdom, 2015.

[26] H. W. Chen, R. D. Zhu, J. X. Zhu, S. T. Wu, A simple method to measure the twist elastic constant of a nematic liquid crystal. Liquid Crystals 2015, 42 (12), 1738-1742.

[27] H. Wang, T. X. Wu, X. Zhu, S.T. Wu, Correlations between liquid crystal director reorientation and optical response time of a homeotropic cell. J. Appl. Phys. 2004, 95 (10), 5502.

[28] Y. A. Nastishin, R. D. Polak, S. V. Shiyanovskii, V. H. Bodnar, O. D. Lavrentovich, Nematic polar anchoring strength measured by electric field techniques. J Appl Phys 1999, 86 (8), 4199-4213.

[29] V. G. Chigrinov, V. M. Kozenkov, H.-S.Kwok, Photoalignment of liquid crystalline materials : physics and applications, John Wiley & Sons: Chichester, 2008.

[30] O. Yaroshchuk, Y. Reznikov, Photoalignment of liquid crystals: basics and current trends. J Mater Chem 2012, 22 (2), 286-300.